\documentclass[12pt]{iopart}
\usepackage[utf8]{inputenc}
\usepackage{graphicx}
\usepackage[usenames]{color}
\usepackage{verbatim}
\usepackage{iopams}

\begin{document}
\title {Atomic wave packet dynamics in finite time-dependent optical lattices}
\author{T Lauber$^1$, P Massignan$^{2,3,4}$, G Birkl$^1$, and A Sanpera$^{5,2,4}$}
\address{$^1$ Technische Universit\"at Darmstadt, Institut f\"ur Angewandte Physik, Schlossgartenstraße 7, D-64289 Darmstadt, Germany.}
\address{$^2$ F\'isica Teorica: Informaci\'o i Processos Qu\`antics, Universitat Aut\`onoma de Barcelona, 08193 Bellaterra, Spain.}
\address{$^3$ ICFO-Institut de Ci\`encies Fot\`oniques, Mediterranean Technology Park, 08860 Castelldefels (Barcelona), Spain.}
\address{$^4$ Kavli Institute of Theoretical Physics, University of California, Santa Barbara, 93106, USA.}
\address{$^5$ ICREA-Instituci\'o Catalana de Recerca i Estudis Avan\c{c}ats, 08010 Barcelona, Spain.}

\ead{thomas.lauber@physik.tu-darmstadt.de}
\begin{abstract}
Atomic wave packets in optical lattices which are both spatially finite and time-dependent exhibit many striking similarities with light pulses in photonic crystals.
We analytically characterize the transmission properties of such a potential geometry for an ideal gas in terms of a position-dependent band structure.
In particular, we find that at specific energies, wave packets at the center of the finite lattice may be enclosed by pairs of band gaps.
These act as mirrors between which the atomic wave packet is reflected, thereby effectively yielding a matter wave cavity.
We show that long trapping times may be obtained in such a resonator and investigate the collapse and revival dynamics of the atomic wave packet by numerical evaluation of the Schrödinger equation.
\end{abstract}
\submitto{\jpb}

\section{Introduction}
Ultracold atoms confined in periodic optical potentials, so called optical lattices, are versatile systems which permit us to address a large variety of open problems in condensed matter physics \cite{PethickSmithBook,Stringari2003,Bloch2005,Lewenstein2007}.
In addition, optical lattices allow us to go even one step beyond what can be realized in usual condensed matter systems. For instance, the optical lattice depth can be suddenly or adiabatically modified in time inducing a change on the wave packet dynamics on time scales much shorter than the typical atomic coherence times. Moreover, optical lattices have become fundamental ingredients in the realization of quantum devices as diverse as atomic clocks \cite{Ye2008}, gravitometers \cite{Ferrari2006}, and interferometers \cite{Cronin2008}.

Atomic quantum devices generally require an extremely precise and coherent control of the atomic motion, the internal states of the atoms, and their interactions. The toolbox available for the manipulation of ultracold gases is vast, and it is nowadays routinely possible to change almost at will the spatial dimensionality of the gas, the temperature, the number and the statistics of the atomic species, and the strength of the interparticle interactions. For controlling the external degrees of freedom, there is a large freedom in the choice of externally confining potentials.
As an example, it is possible to guide neutral atoms by means of atomic waveguides. These can be created either optically, using tightly focused laser beams \cite{Bongs2001} or micro-fabricated optical elements \cite{Birkl2001,Dumke2002} and magnetically, by electric circuits imprinted on atom chips \cite{Hinds1999,Dekker2000, Folman2002, Reichel2002, Fortagh2007}. Both schemes generate an effectively one-dimensional guide, which strongly suppresses the motion of the atoms in the transversal directions.

\begin{figure}
\begin{center}
\includegraphics[width=0.65\textwidth]{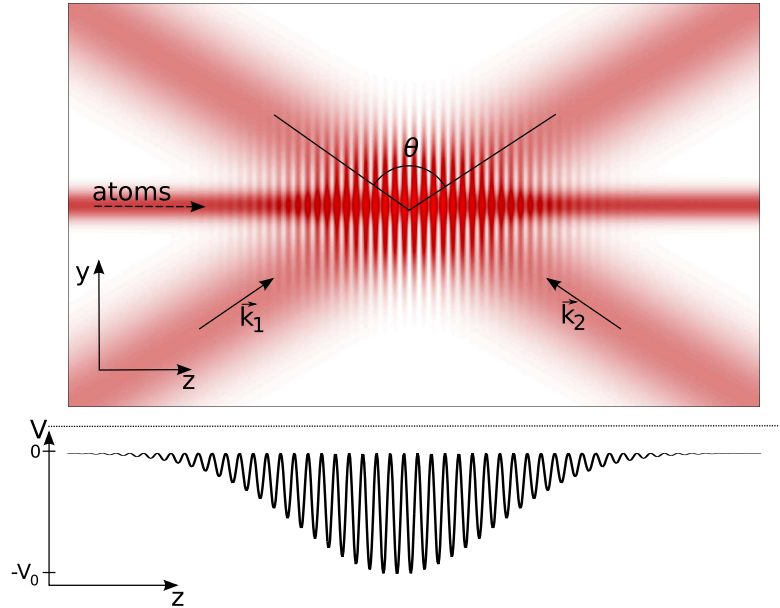}
\caption{Sketch of the proposed setup. Two laser beams with wave vectors $\vec{k_1}$ and $\vec{k_2}$ are intersected at an angle $\theta$ to form a finite optical lattice, while an extra beam propagating along the $\vec{z}$ direction creates a waveguide for the atoms. The lower part shows the potential resulting from the finite lattice along the waveguide axis.}
\label{figpotential}
\end{center}
\end{figure}

Some quantum devices may require specific shaping or filtering of an atomic wave packet. A promising potential structure to shape a wave packet is a 1D optical lattice with spatially finite extent. The propagation of the wave packet in such a structure mimicks closely the behavior of photons in a (1D) photonic crystal
\cite{Yablonovitch1994,Joannopoulos1997}.
Previous studies of atomic wave packets in finite lattices show that this potential may be used as a momentum-selective filter \cite{Santos1997,Santos1998}, or as a matter wave cavity capable of producing multiple wave packet reflections with pulsed emission \cite{Santos1999}.
 All the above phenomena appear for attractive as well as for repulsive potentials, yielding both classical and quantum trapping \cite{Carusotto2000a,Friedman1998}. In the presence of atom-atom interactions, the finite periodic structure gives rise to the occurence of non-linear effects, such as the creation of solitons \cite{Carusotto2002}.

In this paper, we develop an analytical model for predicting the main properties of an ideal gas propagating in a finite optical lattice. Our results agree well with those obtained by numerical integration of the Schr\"odinger equation. After analyzing the filtering and shaping capabilities of a stationary potential, we extend our studies to time-varying optical potentials, showing how the spatial variation of the band gaps may be exploited to create a matter wave cavity, in which the wave packet can be trapped and experiences characteristic collapses and revivals. 

\section{Setup}

A periodic potential with finite envelope can be realized by intersecting two identical, not counter-propagating laser beams, as sketched in Fig.~\ref{figpotential}.
We consider two Gaussian laser beams, which are intersecting at angle $\theta$, each with beam waist $w_b$, wave number $k_b$, and individually giving a potential with depth $V_0/4$. The resulting lattice has an envelope given by the transverse intensity distribution of each beam and the angle $\theta$.
 
We add a one-dimensional waveguide to transversely confine the atomic wave packet. The waveguide may be realized by a red-detuned focused laser beam with waist $w_g$ and potential depth $V_g$, propagating along the z-axis. The laser beam producing the waveguide has to be either far-detuned or perpendicularly polarized with respect to the lattice beams to prevent interference effects. We require that the atomic wave packet is in the transversal ground state of the waveguide, i.e., that the total energy $E$ of the wave packet is small compared to a single quantum of the transversal harmonic oscillator, $E<\hbar \omega_\mathrm{ho}$. Here $\omega_\mathrm{ho}=\sqrt{4V_g/mw_g^2}$, with $m$ being the mass of the atom. This condition ensures that the dynamics is effectively limited to the longitudinal direction only, and that the wave packet has a transverse width given by the size of the harmonic oscillator ground state, $a_\mathrm{ho}=\sqrt{\hbar/m\omega_\mathrm{ho}}$. The condition on the energy sets a tight bound on the initial temperature of the sample, but for higher temperatures the calculation can be extended in a straightforward fashion to accomodate excited transverse eigenstates \cite{Kreutzmann2004}. Although the phenomena presented below are quite general, in this paper we consider the experimentally appealing situation of $^{87}$Rb atoms guided in dipole potentials far red-detuned to the $5S_{1/2}\rightarrow 5P_{3/2}$ transition.

 Close to the axis of the waveguide ($x^2+y^2 < a_\mathrm{ho}^2$), the potential created by the combination of the finite lattice and the waveguide is
\begin{eqnarray}
V_\mathrm{3D}(x,y,z)=&-V_0\exp \left(-\frac{2(x^2+y^2)}{w_\perp ^2}-\frac{2z^2}{w_z ^2}  \right)\cos ^2\left(k_L z \right)\nonumber\\
&-V_g\exp \left(-\frac{2(x^2+y^2)}{w_g ^2} \right)
\end{eqnarray}
Here, $w_z=w_b/\cos\left( \theta /2 \right)$ and  $w_\perp=w_b/\sin\left( \theta /2 \right)$ are the waists of the finite lattice along the axis of the waveguide and transversally to it, respectively. The wave number $k_L=k_b\cdot\sin\left( \theta/2\right)$ of the lattice depends on the angle between the beams, and on the wave number of laser beams creating the lattice. 
In the following, all momenta and energies will be expressed in terms of the lattice recoil units, $p_R=\hbar k_L$ and $E_R=p_R^2/2m$.

Under the assumptions that atoms are trapped on the axis of the waveguide, and $w_g \ll w_\perp$ and $V_g \gg V_0$ the variation of the lattice potential in the transverse direction can be neglected. Therefore, one is left with a 1D optical lattice potential,
\begin{eqnarray}
V_\mathrm{1D}(z)=-V_0\exp \left(-\frac{2z^2}{w_z ^2} \right) \cos ^2\left(k_L z \right)
\end{eqnarray}
which has a Gaussian envelope with width $w_z$.
 
We assume that the atomic wave packet is initially located inside an additional tightly confining potential well $V_{in}$ along $z$ at a large distance $z_0$ from the finite lattice structure, e.g. at $z_0 < -3w_z$. The wave packet is released from this well and is accelerated at the beginning of each realization. This can be done, for example, by applying a Bragg pulse \cite{Kozuma1999}, hence adding an initial momentum $p_{\mathrm{in}}$ to the wave packet. 
 This sets the expanding wave packet in motion towards the finite lattice, in analogy with a light pulse impinging on a photonic crystal.

Interactions between dilute ultracold bosons may be characterized in terms of a two-body contact potential whose strength is proportional to the s-wave scattering length $a$. For sufficiently large densities, interactions play an important role in the dynamics of a trapped ultracold gas, and upon release from a trapping potential, these cause a rapid expansion of the gas. As a consequence, typically already a few milliseconds after release, the gas becomes so dilute that the condition $na^3\ll 1$ is satisfied, and interactions may be safely neglected in the dynamics which follow the rapid expansion. Interaction effects in finite optical lattices, and in particular the formation of bandgap solitons, have been discussed previously in ref.\ \cite{Carusotto2002}. In the following, we consider the dynamics of a wave packet in a waveguide after a sufficiently long time after the release from $V_{in}$. Since we are interested in the dynamics over time scales which are long compared to the early interaction-driven expansion, we can, without loss of generality, assume that each particle behaves independently.

The Hamiltonian describing the dynamics along the z-axis therefore adopts the simple form
\begin{equation}
\label{Hamiltonian}
H=-\frac{\hbar^2\partial^2_z}{2m}-V_0\exp \left(-\frac{2z^2}{w_z ^2} \right) \cos ^2\left(k_L z \right).
\end{equation}

\section{Finite optical lattices as momentum filters}

\begin{figure}
\begin{center}
\includegraphics[width=\textwidth]{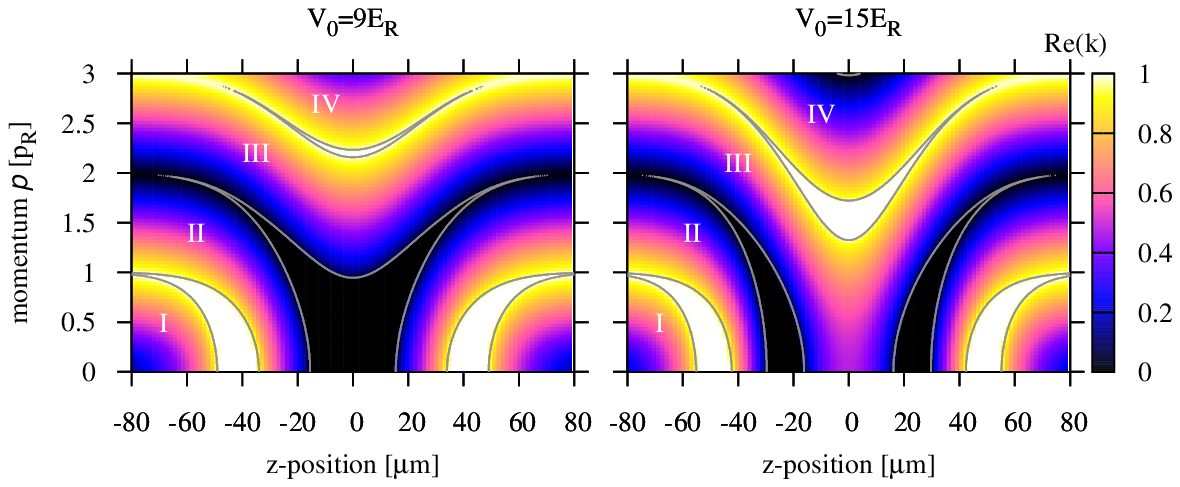}
\includegraphics[width=\textwidth]{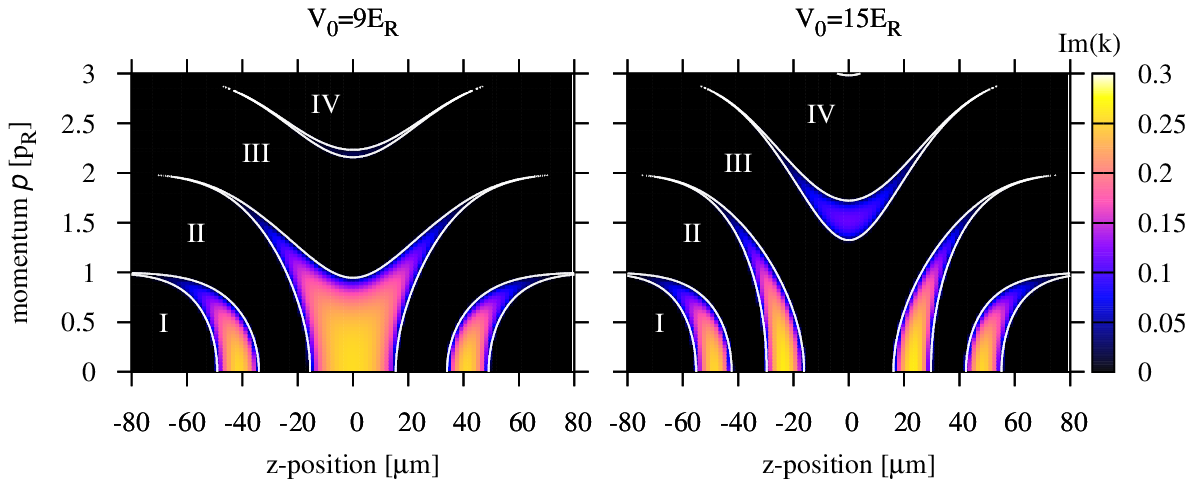}
\caption{Position-dependent band structure of a finite optical lattice with $w_z=50\mu m$. The upper and lower parts depict $|\mathrm{Re}[k(z,p)]|/k_R$ and $\mathrm{Im}[k(z,p]/k_R$, respectively, for two different values $V_0$ of the maximum optical lattice depth. The roman numbers index the different bands and the continuous lines mark the band edges.}
\label{figimk}
\end{center}
\end{figure}
\begin{figure}
\begin{center}
\includegraphics[width=0.6\textwidth]{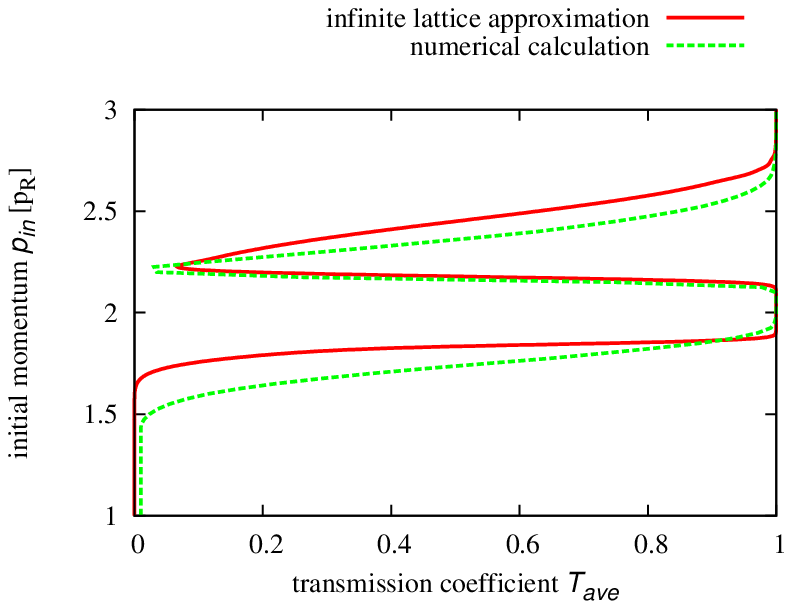}
\caption{Transmission of a Gaussian wave packet through a finite optical lattice. As a function of the average momentum $p_{\mathrm{in}}$ of the incoming wave packet, we plot the transmission coefficient as obtained by \Eref{reflectAnalytical} (solid, red), and by direct integration of the Schr\"odinger equation (dashed, green). Here $V_0=9E_R$, $\sigma_p=0.0325p_R$, and $w_z=50\mu m$.
The momentum axis is plotted vertically for easier comparison with \Fref{figimk}.
}
\label{figreflectivity}
\end{center}
\end{figure}

In this section, we consider the transmission properties of a finite optical lattice, showing that a simple analytical model is able to reproduce in detail the transmission and reflection for different incoming momenta. In the limit of $P\ll \Delta x \ll w_z$ where the spatial extent of the wave packet $\Delta x$ is large compared to the lattice period $P=\lambda_L/2=\lambda_b/(2\sin(\Theta/2))$, but small compared to the waist of the lattice $w_z$, one may assume that at each point along the propagation axis, the wave packet experiences a periodic potential with the local depth but with infinite extension \cite{Santos1997,Santos1998}. This approximation, the so called "infinite lattice" approximation, is equivalent to the one commonly employed in geometrical optics to describe propagation in non-uniform media. For electromagnetic waves the more complex non-adiabatic phenomena due to linear coupling between the different polarization components can lead to additional modifications \cite{Zheleznyakov1983}. Within the "infinite lattice" approximation, the Hamiltonian in  Eq.~(\ref{Hamiltonian}) may be directly diagonalized at each point, yielding the position-dependent band structure shown in \Fref{figimk}.
When the energy of the wave packet lies inside a bandgap (coloured regions in the lower panels of Fig.~\ref{figimk}), the quasimomentum acquires an imaginary part, which describes an exponential attenuation of a traveling wave due to the presence of a band gap. At the same time, the real part is fixed at the center (e.g. $Re(k)=0$) or the edge ($Re(k)=1$) of the Brillouin zone.
The transmission coefficient for a wave packet crossing the structure from $z=-\infty$ to $z=+\infty$ may be calculated analytically 
by integrating the imaginary part of the wave vector along the propagation direction. If all components of a wave packet belong to an allowed band, the transmittivity through the structure approaches unity. If instead some components enter a band gap along their propagation through the structure, part of the wave packet will be reflected backwards. For a given momentum, the wave packet may encounter one or multiple gaps. 

Since we consider an optical potential with a Gaussian envelope which is symmetric around the point $z=0$, the band structure is also symmetric with respect to the same point. For a monochromatic wave with momentum $p$, the transmission coefficient for the particle density is given by $T(p)=T^2_+(p)$, where
\begin{equation}
T_+(p)=\exp\left( -2\int_{0}^\infty \mathrm{d}z\,  \mathrm{Im}[k(z,p)] \right) 
\end{equation}
is the transmission coefficient for the wave propagating from the center of the lattice ($z=0$) to $z=+\infty$.
Whenever a wave packet is surrounded by a pair of gaps, those behave as mirrors between which the atoms may be multiply reflected. For simplicity, we will consider here only the case of a wave packet which experiences at most one pair of band gaps, but the present argument may be straightforwardly generalized to more complex situations. If the point $z=0$ belongs to an allowed band, to calculate the transmission coefficient, one has to take into account the possibility of multiple reflections between two gaps. The total transmission coefficient for a given momentum allowing for multiple reflections is given by
\begin{eqnarray}
\label{TMR}
T(p)&=&T_+^2(p)+ T_+(p)(1-T_+(p))^2 T_+(p)+ T_+(p) (1-T_+(p))^4 T_+(p) +\ldots\\
&=&T_+^2(p)\cdot\sum\limits_{n=0}^\infty(1-T_+(p))^{2n}=\frac{T_+(p)}{2-T_+(p)}\nonumber
\end{eqnarray}
The various contributions on the right hand side of \Eref{TMR} describe respective trajectories through the structure with 0,2,4,\ldots reflections inside the cavity.

This result gives the transmission coefficient for a monochromatic wave. If we consider a Gaussian wave packet with a spread in momentum $\sigma_p$ around the average momentum $p_\mathrm{in}$, the latter expression has to be convoluted with the momentum distribution yielding
\begin{eqnarray}
T_\mathrm{ave}(p_\mathrm{in})=\frac{1}{\sqrt{\pi}\sigma_p }\int_0^{\infty}\mathrm{d}p\,T(p)\cdot\exp\left( -\frac{(p-p_\mathrm{in})^2}{\sigma_p^2}\right)
\label{reflectAnalytical}
\end{eqnarray}

As shown in \Fref{figreflectivity}, the analytic result (solid) reproduces the main features of the full simulation of the time evolution of the wave packet obtained by integrating the Schr\"odinger equation (dashed). In the simulations the lattice period is chosen to be $P=390$nm and the Gaussian envelope waist is $w_z=50\mu$m. This can be achieved for example by crossing two laser beams with wavelength $\lambda_b=1064$nm and waist $w_b=34\mu$m under an angle of $\Theta=94^\circ$.

 Discrepancies due to finite size effects appear at the regions where the spatial extent of the band gaps gets comparable to the size of the wave packet. There, the "infinite lattice" approximation is poorly verified.

\section{Time dependent potentials: Trapping atoms in a band gap cavity}
The analytical band-structure calculation above gives us a tool to design flexible potentials with specific characteristics. These could be, for example, the analog to a photonic crystal with transmission windows at given energies \cite{Yablonovitch1994,Joannopoulos1997}. One is also able to estimate at which repetition rate and intensity several matter wave pulses are emitted from the cavity in the case where multiple reflections occur.

Even more possibilities are opened by changing the finite potential with time. A possibility is the introduction of a time-dependent phase shift between the two lattice beams, which yields a moving lattice. This results in a shift in momentum of the whole band structure. Also the angle between the laser beams could be varied in time, with the effect of dynamically changing the lattice constant \cite{Kasamatsu2008,Li2008}.

\begin{figure}
\begin{center}
\includegraphics[width=\textwidth]{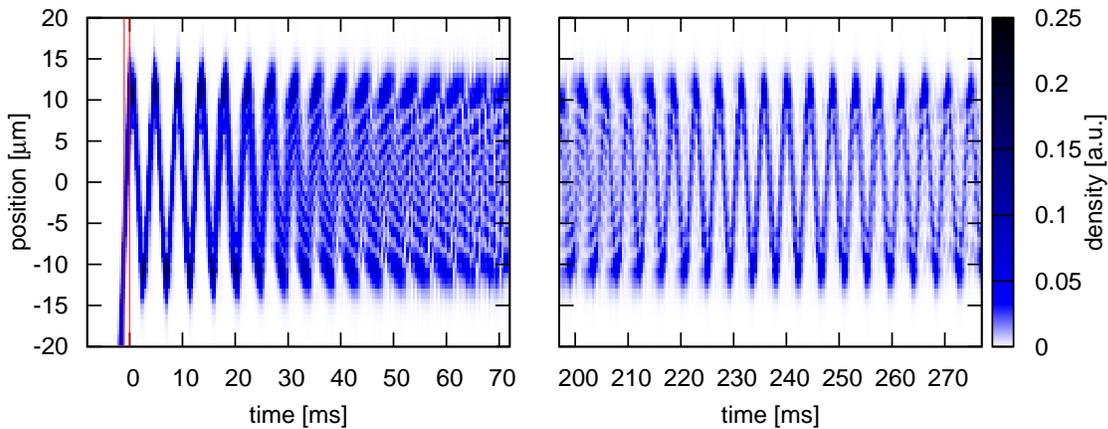}
\caption{Revival dynamics of a wave packet with initial momentum $p_\mathrm{in}=2.4p_R$ and momentum spread $\sigma_p=0.0325p_R$ trapped in the finite lattice. The peak lattice depth is ramped linearly from $9E_R$ to $15E_R$ in $1$ms between the vertical lines around $t=0$. This sequence creates an almost perfect cavity, with very limited loss. The wave packet oscillates inside the cavity, spreads within the first 40 ms, but after 240 ms the dynamics show a clear revival.}
\label{figrevival}
\end{center}
\end{figure}

Alternatively, one may change the intensity of the lattice beams to vary the depth $V_0$ of the potential. This induces a bending and shifting of the band gap regions as shown in \Fref{figimk} ($V_0=9E_R$ vs. $V_0=15E_R$).
If the lattice depth is increased adiabatically, the quasimomentum of the atoms inside the lattice is conserved, implying that the atoms do not change band but simply follow the band and shift to lower energies. Adiabaticity requires $\frac{d\omega}{dt}\ll\omega^2$ \cite{Ketterle1999} which is, for the high trapping frequencies $\omega$ achieved in the axial direction of a lattice, easily fulfilled. For parameters used in the example of \Fref{figrevival} one obtains $t_\mathrm{ramp}\gg5\mu s$. On the other hand, the atoms should not travel too far during $t_\mathrm{ramp}$ so that one is able to change the potential during the time the atoms spend between two gap locations. For the parameters considered here, rubidium atoms with $2\hbar k_L$ momentum travel at a velocity of approximately 10$\mu$m per millisecond which give a time $t_\mathrm{ramp}\leq 2$ms. Given these  constraints, we usually chose a time of 1ms for the linear ramp of the potential depth for our simulations.

An increase in lattice depth yields more reflective band gaps, as seen by the increase of the corresponding imaginary part of the wave vector in \Fref{figimk} ($V_0=9E_R$ vs. $V_0=15E_R$).
By changing the lattice depth when the wave packet is at the center of the lattice between two gaps, one is able to trap the atoms in a configuration reminiscent of a cavity closed by two Bragg-like mirrors. By choosing the parameters carefully, reflectivities close to unity may be achieved. In order to remain in the quantum trapping regime, the change in potential energy has to be smaller than the kinetic energy, so that the total energy is still positive and the atoms are not trapped classically.

Exploiting time-dependent potentials, one can easily realize long trapping times. As an example, we show in  \Fref{figrevival} the result of a numerical integration of the Schrödinger equation yielding a lifetime larger than 300ms in the cavity. Every time the wave packet hits a mirror (i.e a band gap) of the cavity, most of the wave packet is reflected but a small fraction is transmitted and escapes the finite lattice. In the initial part of the dynamics (up to 40ms), when the wave packet width is small compared to the cavity length, one observes separate transmitted wave packets and achieves the analog to a pulsed atom laser. Subsequently the wave packet spreads over the whole cavity, but at specific times (e.g. $t\approx 240$ms) the pulsed emission is recovered as the wave packet cyclically re-acquires its original shape.

\section{Collapse and revival dynamics}\label{secdynamics}
The dynamics of a wave packet inside a finite optical lattice exhibit interesting collapse and revival dynamics, as seen in \Fref{figrevival}. Due to the finite momentum width $\sigma_p$, the trapped wave packet rapidly dephases (collapse) and fills the cavity, but at specific times, we observe a clear compression of the atom pulse back to its original shape (revival). We investigated the time needed for such revivals to occur. In order to get a detailed insight on the problem, we first analyze the dynamics in a box potential, and then address the question of whether the same argument may be extended to the band gap cavity studied here.

\subsection{Revivals in a box potential}
In a square well with infinite walls, the revival dynamics is exact and its periodicity may be calculated analytically due to the simple scaling of the eigenvalues. Indeed, the eigenfunctions of a 1D square well of length $L$ are sine functions with wavelengths $\lambda_n=2L/n$ $(n=1,2,\ldots)$. The associated eigenfrequencies are given by $\omega_n=\hbar k_n^2/2m=\hbar\pi^2n^2/2mL^2$. Therefore, each Fourier component is characterized by a period $T_n=2\pi/\omega_n=4mL^2/\pi\hbar n^2$. Since $T_n=T_1/n^2$, a full rephasing of the wave packet is obtained every 
\begin{eqnarray}
T_\mathrm{rev}=T_1=\frac{4m}{\pi\hbar}L^2. 
\label{trev}
\end{eqnarray}
At half of this period, the even frequency components are rephased, while all the odd frequency components are out-of phase by $\pi$. At $T_\mathrm{spec}=T_1/2$ one therefore obtains a specular revival, i.e., the wave packet is recomposed in two components with opposite direction of motion.
If, on the other hand, the initial conditions are chosen to be symmetric with respect to the center of the well, the components with even values of $n$ have zero weight, and a full revival is obtained on the shorter timescale $T_\mathrm{sym}=T_1/8$.
In all cases, the revival time scales quadratically with the length of the well $L$. For a detailed investigation of related phenomena (the so-called "quantum carpets") see Ref.\ \cite{Schleich1999}.

\subsection{Revivals in a finite optical lattice}
\begin{figure}
\begin{center}
\includegraphics[width=0.7\textwidth]{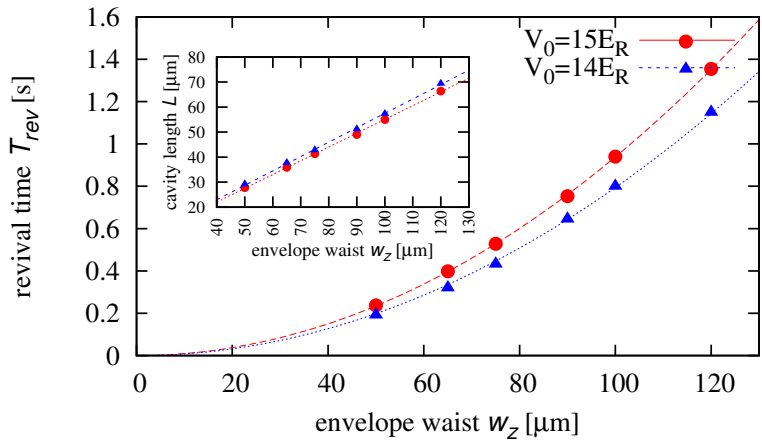}
\caption{Revival time as a function of the waist of the Gaussian envelope $w_z$.
The thin lines are quadratic fits to the numerical data.
 The inset shows the cavity length as a function of the envelope waist.}
\label{figrevivaltime}
\end{center}
\end{figure}

Inside a finite optical lattice cavity, the eigenenergies do not follow the simple scaling found above, and the revivals we observe are only approximate since they are only due to partial constructive interference. Nonetheless, we observe also in this case that the revival time scales quadratically with the waist $w_z$ of the Gaussian envelope, as can be seen in \Fref{figrevivaltime}. In the inset of the same figure, we show that the cavity length $L$, extracted from our simulations as the distance between the two positions where the wave packet is reflected at a gap, scales linearly with the waist $w_z$. Therefore, we find that also for a finite optical lattice $T_\mathrm{rev}\propto L^2$, indicating that a similar description of the revival phenomena as in a box potential can be applied. 

In order to pursue the analogy further, we use the fact that the presence of an infinite periodic potential affects the motion of a particle in a simple manner: it renormalizes its mass. The effective mass of a particle in the n$^{\mathrm{th}}$ band is given by the formula $m^*_n=\hbar^2(\partial^2 E_n/\partial k^2)^{-1}$, where $E_n$ is the energy of the band under consideration. The dynamics of a particle moving in an infinite optical lattice may be effectively described in terms of a free particle, once the bare mass of the particle is replaced by the effective mass $m^*$ \cite{Massignan2003}. Under the "infinite lattice" approximation, a particle bouncing back and forth in a lattice cavity may therefore be considered as a free particle with mass $m^*$ bouncing in a box potential. Considering that the lattice depth changes along the path of the wave packet, we average the effective mass inside the cavity. Hereby, we use a weighing factor given by the inverse of the group velocity in order to take into account the different durations, that the wave packet spends at different positions inside the lattice.
 
\begin{figure}
\begin{center}
\includegraphics[width=0.7\textwidth]{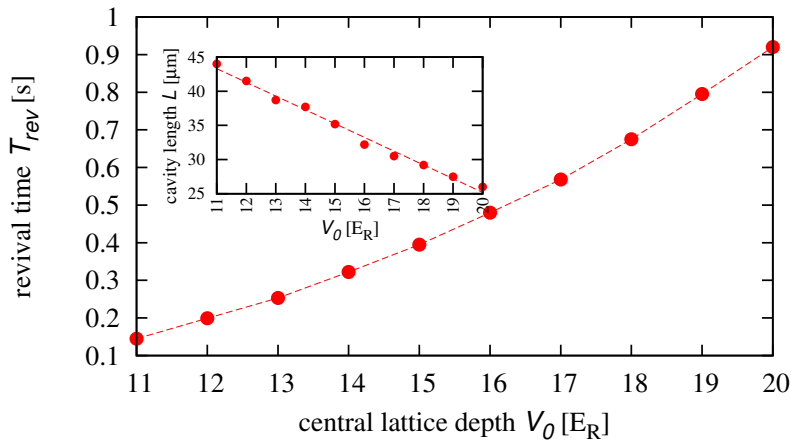}
\caption{Revival time as a function of the central lattice depth $V_0$ for a fixed waist $w_z=65\mu$m. The inset shows the cavity length as a function of $V_0$ of the finite potential. The dashed lines are guides to the eye.}
\label{figrevivaltimeVsDepth}
\end{center}
\end{figure}

Revival times and cavity lengths obtained by numerical simulation are shown in \Fref{figrevivaltimeVsDepth} as a function of the maximum depth of the potential $V_0$. There one observes an increase of the revival time, while the cavity length decreases with larger values of $V_0$ which contradicts Equation \ref{trev} for constant mass.

Following Equation \ref{trev}, the revival time $T_\mathrm{rev}$ can also be calculated using the analytic results: $L$ and $m^*$ may be computed straightforwardly from the band structure shown in \Fref{figimk}.
Although the revival time calculated with the latter method agrees in order of magnitude with the one extracted from numerical integration, a thorough quantitative analysis shows that the analogy with a simple square well can not be achieved quantitatively.
This poor agreement between analytic approximation and full numerical simulation may be caused by finite size effects not included, by the rapid changes of the finite lattice in position space, or by the complicated spatial dependence of the effective mass.
A more detailed analysis of the revival time remains an open and interesting problem, which could be addressed in a future work.
 
\section{Conclusions}
We have studied the properties of an ideal atomic wave packet, which may be generated by releasing a BEC and waiting for the interactions to get negligible, in presence of a finite and time-dependent optical lattice. In analogy with a photonic crystal, the structure behaves as a momentum-selective filter. We have demonstrated how its transmission properties are well captured by a theoretical model based on a spatially-dependent spectrum. By varying the intensity of the lattice beams and thus the potential depth in time, we have shown how a finite lattice can be used as a matter wave cavity (band gap cavity) with long lifetime for atomic wave packets. We have also investigated the collapse and revival dynamics of a trapped wave packet by means of the Schrödinger equation, and we have shown that the revival time scales quadratically with the cavity length, similar to a square well potential.

Possible applications of the configuration described in this paper include the realization of an atom laser with pulsed emission into a waveguide, or the reversible storage and release of a wave packet. The device could also be used as a wave packet splitter, since at half of the revival time the cavity contains two identical counterpropagating wave packets. If the lattice is switched off adiabatically at this moment, two wave packets with opposite momentum are generated.

The experimental techniques needed to realize such potentials are readily available, and we expect that the present paper will stimulate further experiments heading towards the implementation of photonic band gap configurations for atomic wave packets.

\ack
We wish to thank Johannes Küber and Riccardo Sapienza for insightful discussion. P. M. and A. S. acknowledge the Kavli Institute of Theoretical Physics, Santa Barbara, California, USA, where part of this work has been performed. This research was supported in part by the National Science Foundation under Grant No. NSF PHY05-51164.  P. M. and A. S. acknowledge the Spanish MEC projects FIS2008-01236, FIS2008-00784, QOIT (Consolider Ingenio 2010) and the Catalan project 2009-SGR-985. T. L. and G. B. acknowledge the financial support by the DAAD (Contract No. 0804149).

\section*{References}

\bibliography{finitelattice}{}
\bibliographystyle{iopart-num}
\end{document}